\documentclass[11pt]{article}

\usepackage{graphicx}
\usepackage{amsmath}
\usepackage{amssymb}
\usepackage{dcolumn}
\usepackage{bm}
\usepackage[utf8]{inputenc}
\usepackage{authblk}

\newcommand{\be}{\begin{equation}}
\newcommand{\ee}{\end{equation}}
\newcommand{\ba}{\begin{eqnarray}}
\newcommand{\ea}{\end{eqnarray}}
\newcommand{\no}{\nonumber \\}
\newcommand{\gsim}{\mathrel{\hbox{\rlap{\lower.55ex \hbox {$\sim$}}
                   \kern-.3em \raise.4ex \hbox{$>$}}}}
\newcommand{\lsim}{\mathrel{\hbox{\rlap{\lower.55ex \hbox {$\sim$}}
                   \kern-.3em \raise.4ex \hbox{$<$}}}}

\def\roughly#1{\mathrel{\raise.3ex\hbox{$#1$\kern-.75em%
\lower1ex\hbox{$\sim$}}}}
\def\lsim{\roughly<}
\def\gsim{\roughly>}

\def\({\left(}
\def\){\right)}
\def\[{\left[}
\def\]{\right]}
\def\<{\langle}
\def\>{\rangle}

\def\l{{\lambda}}

\def\d{{\delta}}
\def\D{{\Delta}}

\def\O{{\Omega}}

\def\a{{\alpha}}
\def\b{{\beta}}
\def\c{{\chi}}

\def\g{{\gamma}}

\def\m{{\mu}}

\def\r{{\rho}}

\def\th{{\theta}}
\def\ph{{\phi}}

\def\x{{\xi}}

\newcommand{\wg}{{\wedge}}

\newcommand{\cN}{{\cal N}}
\newcommand{\anom}{{\text{anom}}}

\title{\bf Holographic Magnetized Chiral Density Wave}

\author[2]{Yanyan Bu\thanks{yybu@hit.edu.cn}}
\author[1]{Shu Lin\thanks{linshu8@mail.sysu.edu.cn}}
\affil[1]{School of Physics and Astronomy, Sun Yat-Sen University, Zhuhai 519082, China}
\affil[2]{Department of Physics, Harbin Institute of Technology, Harbin 150001, China}

\date{\today}

\begin{document}

\maketitle
%\vspace{0.1in}

\begin{abstract}
We explore the end point of the helical instability in finite density, finite magnetic field background discussed by Kharzeev and Yee \cite{Kharzeev:2011rw}. The nonlinear solution is obtained and identified with the (magnetized) chiral density wave phase in literature. We find there are two branches of solutions, which match with the two unstable modes in \cite{Kharzeev:2011rw}. At large chemical potential and magnetic field, the magnetized chiral density wave can be thermodynamically preferred over chirally symmetric phase and chiral symmetry breaking phase. Interestingly, we find an exotic state with vanishing chemical potential at large magnetic field. We also attempt to clarify the role of anomalous charge in holographic model.
\end{abstract}

\newpage

\section{Introduction}

The ground state of hot and dense QCD matter is one of the key questions in the physics of heavy ion collisions and that of neutron star. In the former case, a strong magnetic field can be produced in off-center collisions. In the latter case, a strong magnetic field is believed to exist in the core of neutron star. Magnetic field is known to modify QCD phases in different ways: In the absence of baryon chemical potential, magnetic field enhances chiral symmetry breaking and reduces critical temperature, known as magnetic catalysis \cite{Klevansky:1989vi,Klimenko:1992ch,Gusynin:1995nb} and inverse magnetic catalysis \cite{Bali:2011qj,Bali:2012zg} respectively. At finite quark chemical potential, the QCD phase diagram becomes much enriched. In particular, a variety of inhomogeneous phases appear, including chiral density wave \cite{Nakano:2004cd}, solitonic modulation \cite{Nickel:2009ke,Nickel:2009wj}, crystalline color superconductor \cite{Alford:2000ze}, quarkyonic spiral \cite{Kojo:2009ha} etc. The quark density is crucial in the formation of these inhomogeneities, see \cite{Buballa:2014tba} for a review. The presence of magnetic field tends to widen the inhomogenous phases, leading to magnetized-chiral density wave \cite{Frolov:2010wn,Tatsumi:2014wka} or magnetized kink \cite{Cao:2016fby}, magnetized quarkyonic chiral spiral \cite{Ferrer:2012zq} etc.

Interestingly, the interplay of quark density and magnetic field can also lead to more new phases. This is realized through axial anomaly: at low temperature, effective model studies found inhomogeneous phases including pion domain wall \cite{Son:2007ny,Eto:2012qd}, chiral magnetic spiral \cite{Basar:2010zd}, chiral soliton lattice \cite{Brauner:2016pko} etc, see also \cite{Miransky:2015ava,Kharzeev:2012ph} for comprehensive reviews.
From the viewpoint of thermodynamics, formation of inhomogeneous phases induces an anomalous charge, which can lower the free energy of the system \cite{Son:2007ny,Brauner:2016pko}. However, the nature of anomalous charge remains a mystery. It is desirable to search for the inhomogeneous phases in other approaches. A number of such studies using holographic models have been carried out \cite{Domokos:2007kt,Ammon:2016szz,Nakamura:2009tf,Ooguri:2010xs,Kim:2010pu,Kharzeev:2011rw,deBoer:2012ij}. In this work, we aim at finding the holographic analog of magnetized chiral density wave. This work is inspired by early work by Kharzeev and Yee \cite{Kharzeev:2011rw}, in which they found an unstable helical mode. We will find the end point of the instability and identify it with magnetized chiral density wave (MCDW) phase. The competition of MCDW and conventional chiral symmetry breaking phase and restored phase reveals novel structure. We will emphasize the role of anomaly and attempt to clarify the nature of anomalous charge.

The paper is organized as follows: In Section \ref{sec_intro}, we give a brief review of the holographic model and the known phase diagram for homogeneous phases \cite{Evans:2010iy}. In Section \ref{sec_mcdw}, we present ansatz for MCDW phase and solve it numerically and obtain its thermodynamics. We discuss the role of anomalous charge in MCDW phase in Section \ref{sec_anom}. We summarize and discuss future perspectives in Section \ref{sec_sum}.

\section{A quick review of the model}\label{sec_intro}

%\subsection{The finite density background}
We use the D3/D7 model for our study. The
background contains $N_c$ D3 branes and $N_f$ D7 branes. In the probe limit $N_f\ll N_c$, the background is simply given by black hole background sourced by D3 branes, with the backreaction of D7 branes suppressed. The D3/D7 model is dual to ${\cal N}=4$ Super Yang-Mills (SYM) field and ${\cal N}=2$ hypermultiplets fields, which transform in adjoint and fundamental
representations of the $SU(N_c)$ group respectively. The model is close to
QCD in the sense that the ${\cal N}=4$ and ${\cal N}=2$ fields can be identified as gluons and quarks respectively. The probe limit is analogous to quenched approximation.
The finite temperature background of D3 branes is given by \cite{Mateos:2006nu}:
\begin{align}\label{d3_metric}
ds^2=-\frac{r_0^2}{2}\frac{f^2}{H}\r^2dt^2+\frac{r_0^2}{2}H\r^2dx^2+\frac{d\r^2}{\r^2}+d\th^2+\sin^2\th d\ph^2+\cos^2\th d\O_3^2.
\end{align}
where
\begin{align}
f=1-\frac{1}{\r^4},\quad H=1+\frac{1}{\r^4}.
\end{align}
We set the AdS radius to $1$. The temperature is given by $T=r_0/\pi$.
We also explicitly factorize $S_5$ into $S_3$ and two
additional angular coordinates $\th$ and $\ph$.
There is also a nontrivial Ramond-Ramond form
\begin{align}\label{f5}
F_5=r_0^4\r^3Hfdt\wg dx_1\wg dx_2\wg dx_3\wg d\r+4\cos^3\th\sin\th d\th\wg d\ph\wg d\O_3.
\end{align}
The D7 branes share the worldvolume coordinates with D3 branes. In
addition, they span the coordinates $x_4-x_7$ parametrized by the $S_3$ coordinates. Their position in $x_8-x_9$ plane can be parametrized by polar coordinate, with radius $\r\sin\th$ and angle $\ph$. The rotational symmetry in the $x_8-x_9$ plane corresponds to $U(1)_R$ symmetry in the field theory.
The D7 branes have an additional $U(1)_B$ symmetry carried by its worldvolume gauge field. In comparison with QCD, the $U(1)_R$ and $U(1)_B$ symmetries are identified as axial and baryon symmetries respectively.

With the background metric \eqref{d3_metric}, the gluons provide a thermal bath at fixed temperature for quarks. The quark chemical potential and magnetic field are turned on by a nonvanishing $A_t(\r)$ and constant $F_{xy}=B$. The phase diagram has been obtained by Evans et al \cite{Evans:2010iy}, showing a rich structure. There is one order parameter of the system, namely chiral condensate. The condensate is determined by the embedding of D7 branes in the D3 brane background. There are two possible embeddings for D7 branes: black hole embedding and Minkowski embedding, corresponding to chirally symmetric ($\c S$) phase and chiral symmetry breaking ($\c SB$) phase. The phases can further be classified based on quark number density. For $\c$S phase, only finite density state is allowed. For $\c$SB phase, both finite density and zero density states are allowed.
In total, three homogeneous phases are found in \cite{Evans:2010iy}, zero density, $\c SB$ phase, finite density, $\c SB$ phase and finite density, $\c S$ phase.

The action of D7 branes is given by a Dirac-Born-Infeld (DBI) term and Wess-Zumino (WZ) term
\begin{align}\label{S_bare}
&S_{D7}=S_{DBI}+S_{WZ}, \no
&S_{DBI}=-N_fT_{D7}\int d^8\x\sqrt{-\text{det}\(g_{ab}+2\pi\a'
  \tilde{F}_{ab}\)}, \no
&S_{WZ}=\frac{1}{2}N_fT_{D7}(2\pi\a')^2\int P[C_4]\wg \tilde{F}\wg \tilde{F}.
\end{align}
Here $T_{D7}$ is the D7 brane tension. $g_{ab}$ and $\tilde{F}_{ab}$ are the induced metric and worldvolume field strength respectively. Defining
\begin{align}
&F_{ab}=2\pi\a'\tilde{F}_{ab}, \no
&\cN=N_fT_{D7}2\pi^2=\frac{N_fN_c\l}{(2\pi)^4},
\end{align}
we can simplify the action to
\begin{align}\label{S_redef}
&S_{DBI}=-\frac{\cN}{2\pi^2}\int d^8\x\sqrt{-\text{det}\(g_{ab}+{F}_{ab}\)}, \no
&S_{WZ}=\frac{1}{4\pi^2}\cN\int P[C_4]\wg F\wg F.
\end{align}
The embedding function $\th$ and worldvolume gauge fields $A_t$ are determined by minimizing the action.
The asymptotic behaviors of $\th$ and $A_t$ are given by
\begin{align}\label{mc}
&  \sin\th=\frac{m}{\r}+\frac{c}{\r^3}+\cdots,
&  A_t=\mu-\frac{n}{\r^2}+\cdots.
\end{align}
The coefficients $m$ and $c$ are related to
the bare quark mass $M_q$ and chiral condensate $\<\bar{\psi}\psi\>$ as \cite{Mateos:2007vn}:
$M_q=\frac{mr_0}{2\pi\a'}$, $\<\bar{\psi}\psi\>=-2\pi\a'\cN c r_0^3$.
The coefficients $\m$ and $n$ are related to the quark chemical potential $\mu_q$ and quark number density $n_q$ as: $\m_q=\frac{mr_0}{2\pi\a'}$, $n_q=2\pi\a'\cN n r_0^3$.
Below we set $r_0=1$. This amounts to working in units of $\pi T$.

For homogeneous phase, the WZ term is not relevant. However, when $B$ and $\m$ are large, the system is found to contain an unstable mode involving simultaneous fluctuations of $x_8$ and $x_9$ \cite{Kharzeev:2011rw}. It is further conjectured that the end point of this instability is helical phase. The presence of the WZ term is essential to the instability. In the next section, we will find the end point of the instability and identify it with MCDW phase known in literature \cite{Tatsumi:2014wka}.

\section{Magnetized Chiral Density Wave}\label{sec_mcdw}

We start with the following ansatz for MCDW
\begin{align}\label{ansatz}
&A_t=A_t(\r),\qquad  \th=\th(\r),\qquad  \ph=k z.
\end{align}
The last two equations in \eqref{ansatz} can be written equivalently as
\begin{align}
x_8+ix_9=e^{i k z}\r\sin\th(\r).
\end{align}
Note that $A_t$ depends on $\r$ only. It gives rise to a homogeneous quark number density.
The fields $x_8$ and $x_9$ form spiral in the direction parallel to the magnetic field. The limit $k\to0$ reduces to the homogeneous case studied before.
In this limit, $x_8=\r\sin\th$ is dual to chiral condensate:
\begin{align}
\bar{\psi}\psi\propto c.
\end{align}
The ansatz \eqref{ansatz} is simply a chiral rotation of chiral condensate along $z$ direction:
\begin{align}
  \bar{\psi}\psi+i\bar{\psi}i\g_5\psi \propto c\(\cos kz+i\sin kz\).
\end{align}
In the presence of non-trivial $\ph$, the dual field theory contains the following interaction term for quarks \cite{Das:2010yw,Hoyos:2011us}.
%It arises from the following interaction term in the field theory action \cite{Hoyos:2011us}
\begin{align}\label{Sint}
  S_I=-m\bar{\psi}e^{i\ph \g_5}\psi.
\end{align}
The interaction term has no analog in QCD. We are interested in the massless limit, where this term vanishes. Therefore the helical phase corresponds to spontaneous breaking of both chiral symmetry and translational symmetry along $z$.
While 1D long range order is known to be washed out by fluctuations in effective models, with the ground state containing only quasi-long range order \cite{Lee:2015bva,Hidaka:2015xza}. In holographic model, the issue is absent because of suppression of fluctuations in large $N_c$ limit.
%Because of nontrivial profile of $\ph$, we should insist $m=0$ for application to QCD.
%\begin{align}
%S=\frac{1}{4}\(1-\c^2\)\sqrt{2+4B^2+\r^4+\frac{1}{\r^4}}\sqrt{}
%\end{align}
%For a given set of parameters $\m$ and $B$, there are four possible solutions corresponding to black hole embedding and Minkowski embedding with and without spirals. For the spirals phases, the momentum of spiral $k$ exists in a finite window. We need to minmize the Gibbs free energy to find out the favored $k$. The phase diagram follows from the competition of the spiral and non-spiral phases.

Plugging the ansatz \eqref{ansatz} into \eqref{S_redef}, we obtain
\begin{align}\label{S_exp}
  &S=\int d^4xd\r({\cal L}_{DBI}+{\cal L}_{WZ}), \no
  &{\cal L}_{DBI}=\cN\frac{-1+\c^2}{4}\sqrt{2+4B^2+1/\r^4+\r^4}\no
&\times  \sqrt{\frac{1}{\r^6+\r^{10}}\(1+\r^4+2k^2\r^2\c^2\)\(2\r^4(1+\r^4)A_t'^2(-1+\c^2)+\(-1+\r^4\)^2\(1-\c^2+\r^2\c'{}^2\)\)}, \no
  &{\cal L}_{WZ}=-\cN B k A_t'(-2\c^2+\c^4).
\end{align}
We have defined $\c=\sin\th$.
Note that the WZ term depends on gauge potential $C_4$. We fix the gauge following \cite{Kharzeev:2011rw},
\begin{align}\label{c4}
C_4=\(\frac{r_0^2}{2}\r^2H\)^2dt\wg dx_1\wg dx_2\wg dx_3-(\cos^4\th-1) d\ph\wg d\O_3.
\end{align}
Other gauge choice has been used in \cite{Hoyos:2011us}. The difference in fact does not alter bulk solutions for MCDW phase because it only causes a constant shift in total action $\D S=\int d^4xd\r BkAt'= \text{Vol}_4Bk\m$. Clearly it affects thermodynamics. Our forthcoming analysis will also support this gauge choice \eqref{c4}.
The equations of motion can be derived as
\begin{align}\label{eom_var}
  &\frac{\d {\cal L}}{\d\c}-\frac{d}{d\r}\(\frac{\d {\cal L}}{\d\c'}\)=0, \no
  &\frac{\d {\cal L}}{\d A_t}-\frac{d}{d\r}\(\frac{\d {\cal L}}{\d A_t'}\)=0.
\end{align}
Since the action depends on $A_t$ only through its derivative, there is a conserved quantity $\frac{\d {\cal L}}{\d A_t'}$. It is identified with quark number density $n$ \cite{Evans:2010iy}. %Explicitly, it contains contribution from both DBI term and WZ term. Note that axial anomaly arises from the WZ term \cite{}. It is tempting to identify the corresponding contribution as anomalous quark number.
%\begin{align}\label{n_anom}
%  n_q^\text{anom}=-Bk\c^2(-2+\c^2).
%\end{align}
Consequently, we can use
\begin{align}
  \frac{\d {\cal L}}{\d A_t'}=n.
\end{align}
Throughout the paper, we focus on finite density solutions. It is known that only black hole embedding can support finite density solutions \cite{Karch:2007br}. %This does not allow us to explore the full phase diagram. Nevertheless, we can still study the competition of chiral symmetry resotred phase, chiral symmetry broken phaseand MCDW phase.
We search for MCDW solution by numerically integrating horizon solution to the boundary. The horizon solution for black hole embedding is obtained analytically as
\begin{align}
  &\c=c_0+c_2\(\r-1\)^2+\cdots,\no
  &A_t'=2a_2(\r-1)+3a_3(\r-1)^2+\cdots,
\end{align}
with $c_0$ and $a_2$ being two independent parameters. We require the field strength $F_{\r t}=A_t'$ vanishes on the horizon.
Higher order coefficients in the expansion are expressible in terms of $c_0$ and $a_2$. We search for numerical solution with fixed $n$, and then scan the parameter $n$. Since $n$ is invariant along the radial direction, we can use $n$ to fix one of the horizon parameter $a_2$:
\begin{align}
  2Bc_0^2k-Bc_0^4k+\frac{a_2\sqrt{1+B^2}\(1-c_0^2\)^2\(1+c_0^2k^2\)}{\sqrt{\(1-a_2^2\)\(1-c_0^2\)\(1+c_0^2k^2\)}}=n.
\end{align}
Note that $\c=\sin\th$, thus $0<c_0<1$. For a given set of parameters $n$, $B$ and $k$, $c_0$ is to be determined by the boundary condition $m=0$. In general, the MCDW solution exists for continuous values of $k$ at large $n$ and $B$. To find out the preferred spiral momentum $k$, we need to minimize Gibbs free energy in grand canonical ensemble. %Note that we cannot minimize Helmholtz free energy in canonical ensemble because as we will see shortly the two ensembles are not equivalent even within the class of MCDW phase.
The quark chemical potential is given by bulk integration of $A_t'$
\begin{align}\label{mu_def}
  \m=\int_1^\infty d\r A_t'.
\end{align}
In practice, we need to tune $n$ and $k$ simultaneously such that $\m$ remains unchanged. This is a numerically challenging task. We are able to achieve $1\%$ percentage accuracy for $\m$.
The Gibbs free energy $\O$ is related to the Euclidean action as
\begin{align}\label{gibbs}
  \O=\frac{1}{\b}S^E=-\int d^3xd\r{\cal L}=-V\int d\r{\cal L}.
\end{align}
The integration of holographic coordinate $\r$ contains divergence. We regularize the action by imposing a UV cutoff $\r=\r_{max}$ and renormalize by adding the following counter terms \cite{Guo:2016nnq}
\begin{align}
  S_{counter}=\r_{max}^4-\frac{m^2\r_{max}^2}{2}+\frac{1}{4}\ln\r_{max}\(2B^2+k^2m^2\).
\end{align}
The appearance of $k$ in the counter term for massive case is not surprising as $k$ appears as a parameter of the theory according to \eqref{Sint}. There is also finite counter term for massive case \cite{Mateos:2007vn}.
The finite counter term does not bother us since we focus on massless case.

The ground state is to be determined by comparing the free energy of the MCDW phase with those of the known $\c$S phase and $\c$SB phase \cite{Evans:2010iy}. The $\c$SB phase appears only at large $B$, while the $\c$S phase exists for any $B$ and finite $\m$. The $\c$SB phase can be obtained as a limit $k\to0$ from the MCDW phase. The $\c$S phase corresponds to the trivial embedding $\c=0$. The free energy is given by the same expression \eqref{gibbs}. To compare the free energy of the three phases, we use the free energy of $\c$S phase as a baseline, i.e. we calculate $\D\O=\O_{\text{MCDW}}-\O_{\c\text{S}}$ for MCDW phase and $\D\O=\O_{\c\text{SB}}-\O_{\c\text{S}}$ for $\c$SB phase.
$\D\O$ of MCDW phase and $\c$SB phase are at percentage level of $\O_{\c\text{S}}$. For the largest magnetic field $B/(\pi T)^2=15$, $\D\O$ is less than $1\%$ of $\O_{\c\text{S}}$, making comparison of free energy more difficult.
%High precision data is needed for study of thermodynamics. We show the competition of phases for $B=6.5$, $B=9$ and $B=15$ in Figures \ref{fig_b65O}, \ref{fig_b9O} and \ref{fig_b15O} respectively.

\begin{figure}[t]
  \centering
\includegraphics[width=10cm]{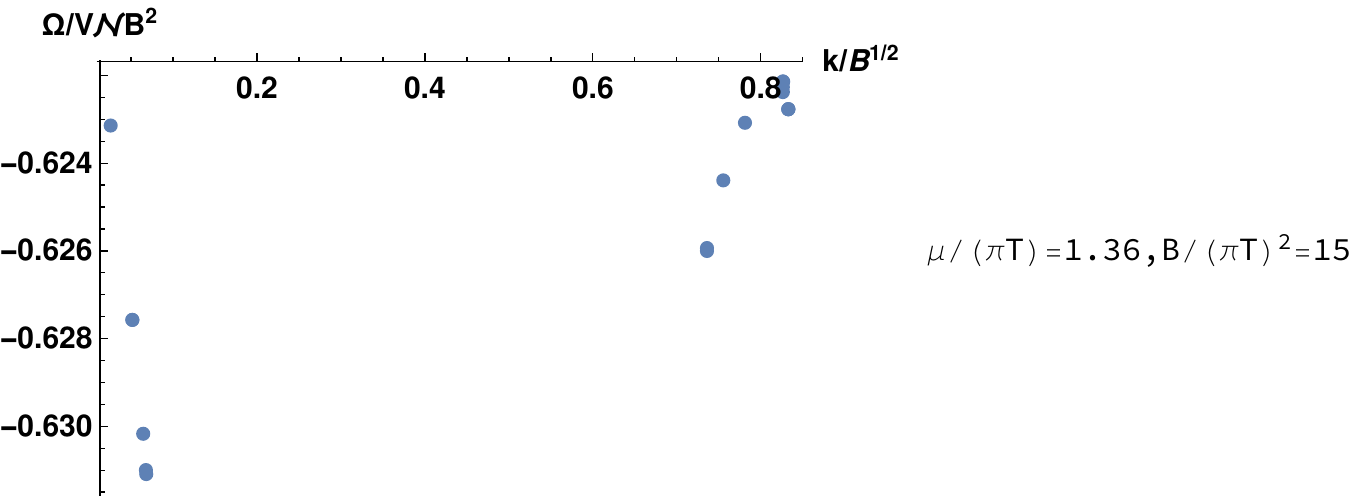}
\caption{\label{two_windows}$\O/\(V\cN B^2\)$ versus $k/B^{1/2}$ at $B/(\pi T)^2=15$ and $\m/(\pi T)=1.36$. Here $\O/V$ is the free energy density with $V=\int d^3x$. The MCDW phase exists in two branches. The lowest free energy is found at the right boundary of the window of smaller $k$.}
\end{figure}

In general, We find MCDW solutions exist in two windows of $k$ at large $\m$ and $B$. The number of windows coincide with the number of unstable modes \cite{Kharzeev:2011rw,Guo:2016dnm} in the chirally symmetric background.
We find the lowest free energy is usually found near the boundary of either window. We show a typical $\D\O$-$k$ plot in Figure \ref{two_windows}.
Although there is only one thermodynamically preferred state, we will keep MCDW states from minimizing free energy in both windows for the purpose of illustration.
Below we present three representative MCDW solutions. They include (i) the case with $B/(\pi T)^2=6.5$, where $\c$SB phase does not exist, and there is competition between $\c$S phase and MCDW phase; (ii) the case with $B/(\pi T)^2=9$, where the large $k$ branch of MCDW phase is thermodynamically preferred in wide region of $\m$; (iii) the case with $B/(\pi T)^2=15$, where the small $k$ branch of MCDW phase is thermodynamically preferred in wide region of $\m$.

\begin{figure}[t]
  \centering
\includegraphics[width=7.5cm]{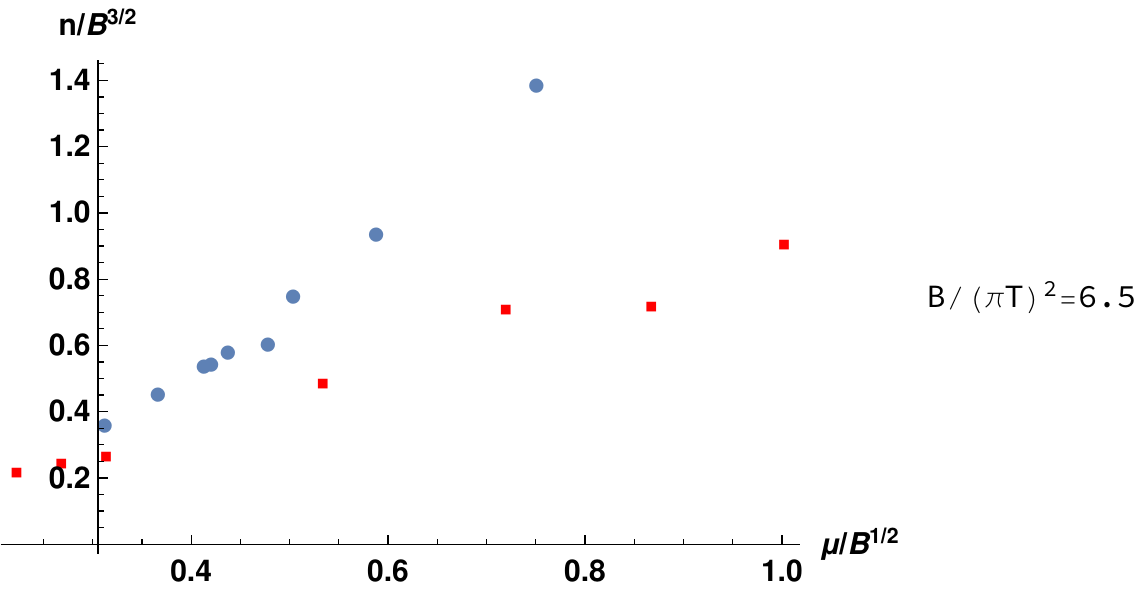}
\includegraphics[width=7.5cm]{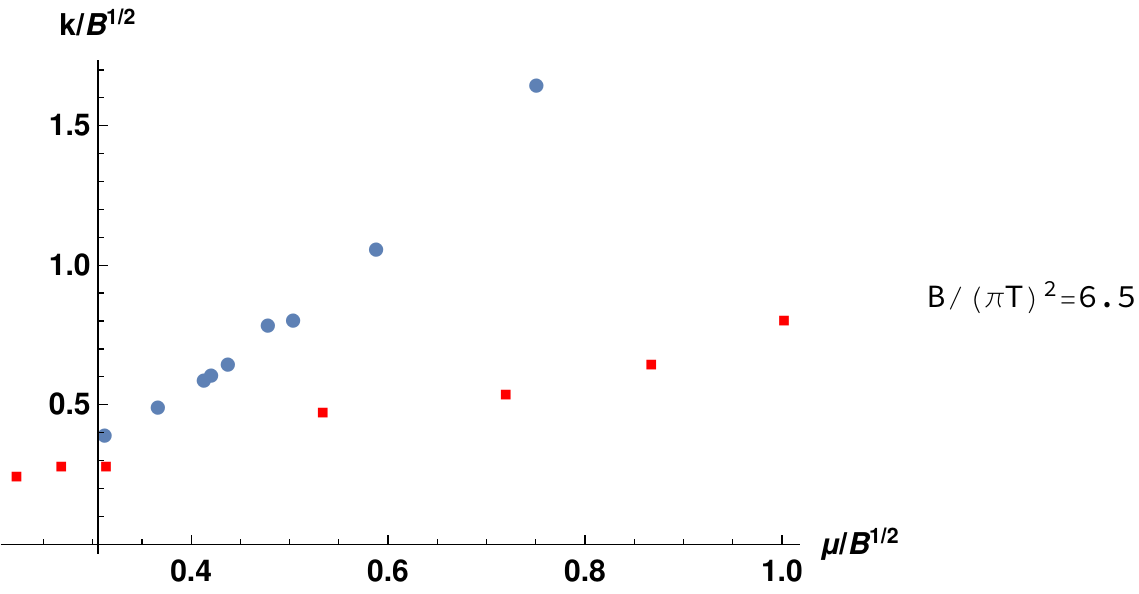}
\caption{\label{fig_b65}$n/B^{3/2}$ versus $\m/B^{1/2}$ (left) and $k/B^{1/2}$ versus $\m/B^{1/2}$ (right) at $B/(\pi T)^2=6.5$. The MCDW phase clearly splits into two branches. The branch with large $k$ and small $k$ are marked by blue disk and red square respectively.}
\end{figure}

\begin{figure}[t]
  \centering
\includegraphics[width=9cm]{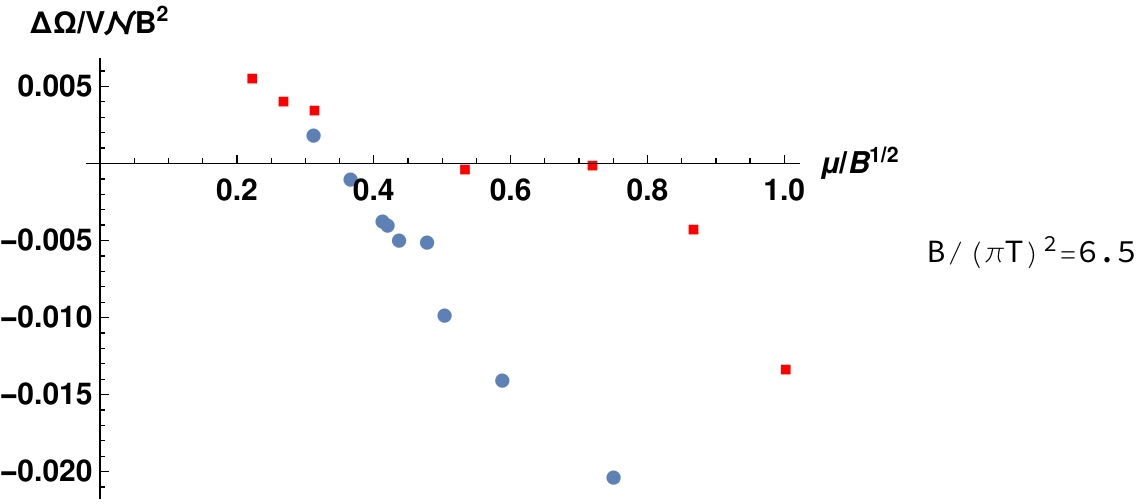}
\caption{\label{fig_b65O}$\D\O/\(V\cN B^{2}\)$ versus $\m/B^{1/2}$ at $B/(\pi T)^2=6.5$ for two branches of MCDW phases, marked by blue disk and red square. The large $k$ MCDW phase has lower free energy than the small $k$ MCDW phase at fixed $\m$. Both are found to have lower free energy than the chirally symmetric phase for large enough $\m$. In particular, the large $k$ MCDW phase becomes thermodynamically preferred above $\m/B^{1/2}\simeq 0.35$. The chiral symmetry breaking phase does not exist at this value of $B$.}
\end{figure}

We first show MCDW phase at $B/(\pi T)^2=6.5$ in Figure \ref{fig_b65}. For a given $\m$, there are two MCDW solutions from the large $k$ branch and small $k$ branch. The large and small $k$ branch of MCDW solution give large and small density $n$ respectively. The corresponding free energy density $\D\O/V$ is shown in Figure \ref{fig_b65O}. At this value of $B$, $\c$SB phase does not exist. There is competition between $\c$S phase and MCDW phase. The large $k$ branch is always thermodynamically more stable than the small $k$ branch, and it dominates over the $\c$S phase when $\m/B^{1/2}\gtrsim 0.35$.

\begin{figure}[t]
  \centering
\includegraphics[width=7.5cm]{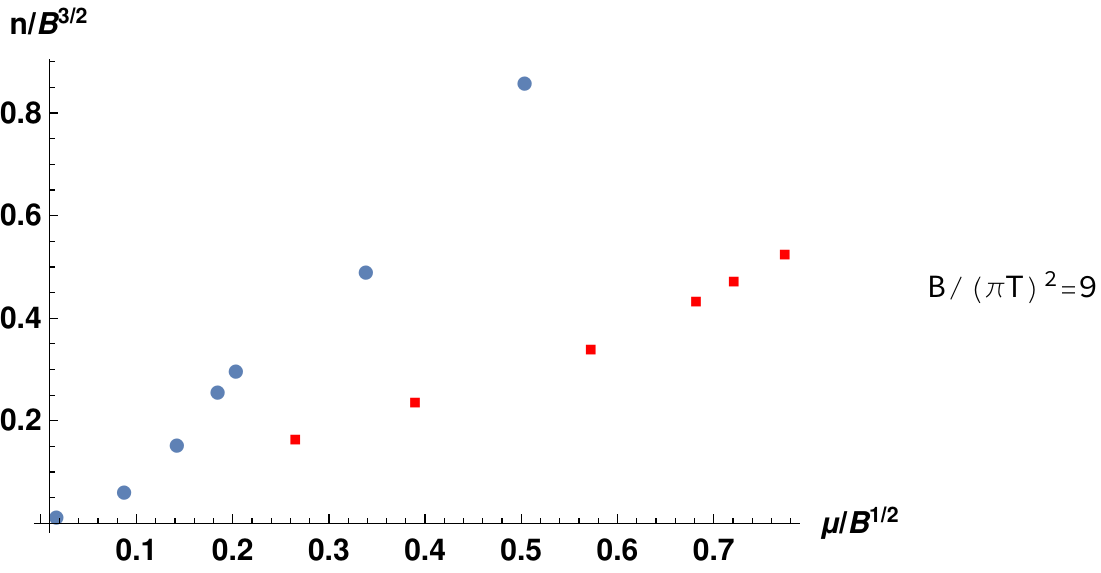}
\includegraphics[width=7.5cm]{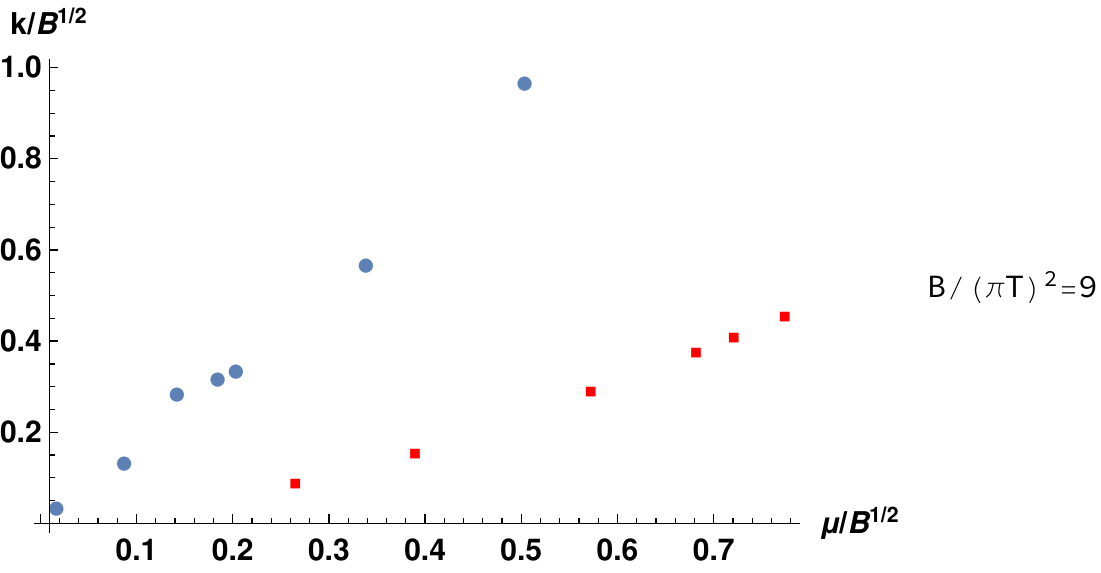}
\caption{\label{fig_b9}$n/B^{3/2}$ versus $\m/B^{1/2}$ (left) and $k/B^{1/2}$ versus $\m/B^{1/2}$ (right) at $B/(\pi T)^2=9$. %The MCDW phase splits into two branches with wider gap in $n$ and $k$.
The branch with large $k$ and small $k$ are marked by disk and square respectively.}
\end{figure}

\begin{figure}[t]
  \centering
\includegraphics[width=9cm]{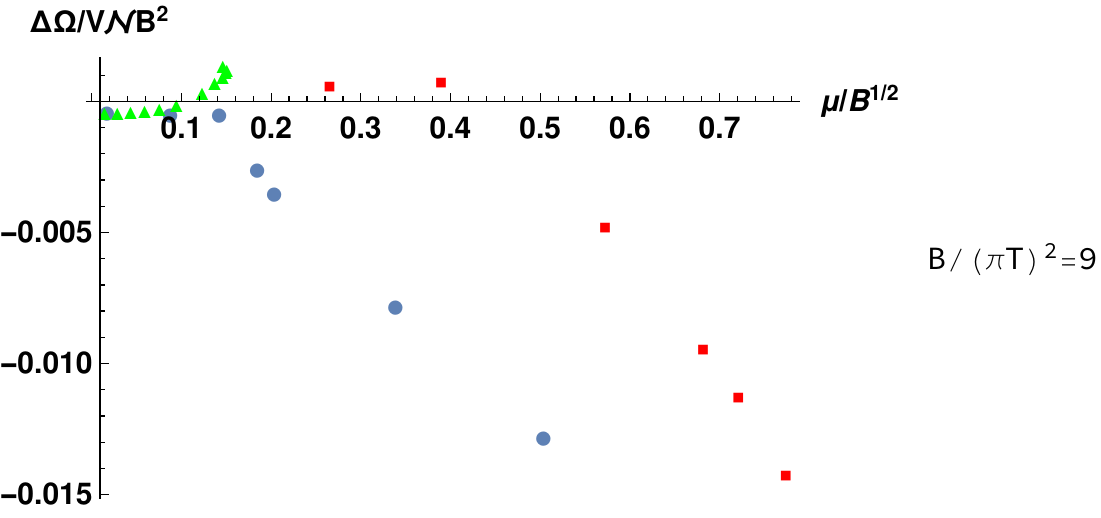}
\caption{\label{fig_b9O}$\D\O/\(V\cN B^{2}\)$ versus $\m/B^{1/2}$ at $B/(\pi T)^2=9$ for two branches of MCDW phases, marked by blue disk and red square and $\c$SB phase marked by green triangle. The large $k$ MCDW phase has lower free energy than chirally symmetric phase and small $k$ MCDW phase in their overlap region. The chiral symmetry breaking case exists below a critical value of $\m/B^{1/2}\simeq 0.15$. Current precision of numerical data does not allow for a decisive conclusion on the preferred state out of MCDW and $\c$SB phase.}
\end{figure}

Next we present the case at $B/(\pi T)^2=9$. In Figure \ref{fig_b9} we show the density and spiral momentum of two branches of solutions. Again the large and small $k$ branch of MCDW solution give large and small density $n$ respectively.
%The gap of $n$ and $k$ between two phases widen compared with the case at $B/(\pi T)^2=6.5$.
The comparison of free energy is shown in Figure \ref{fig_b9O}. We find the MCDW phase with large $k$ is always preferred over $\c$S phase. At low $\m$, $\c$SB phase can occur. Whether $\c$SB phase can be preferred over MCDW phase cannot be decisively answered by current precision of numerical data. Nevertheless, the existence of $\c$SB phase would be constrained in a narrow window of $\m$ if it exists as a thermodynamically preferred state.

\begin{figure}[t]
  \centering
\includegraphics[width=7.5cm]{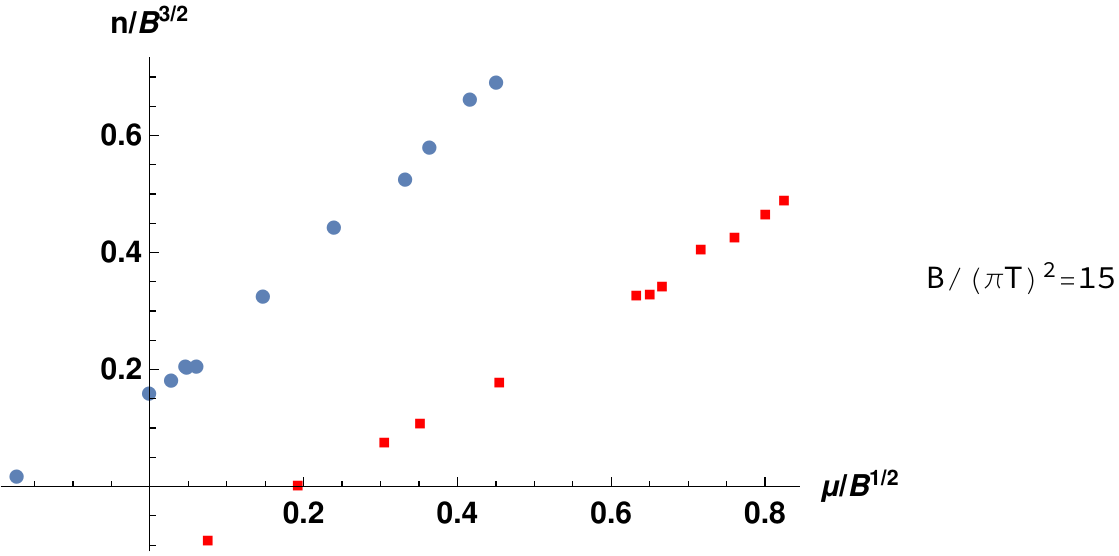}
\includegraphics[width=7.5cm]{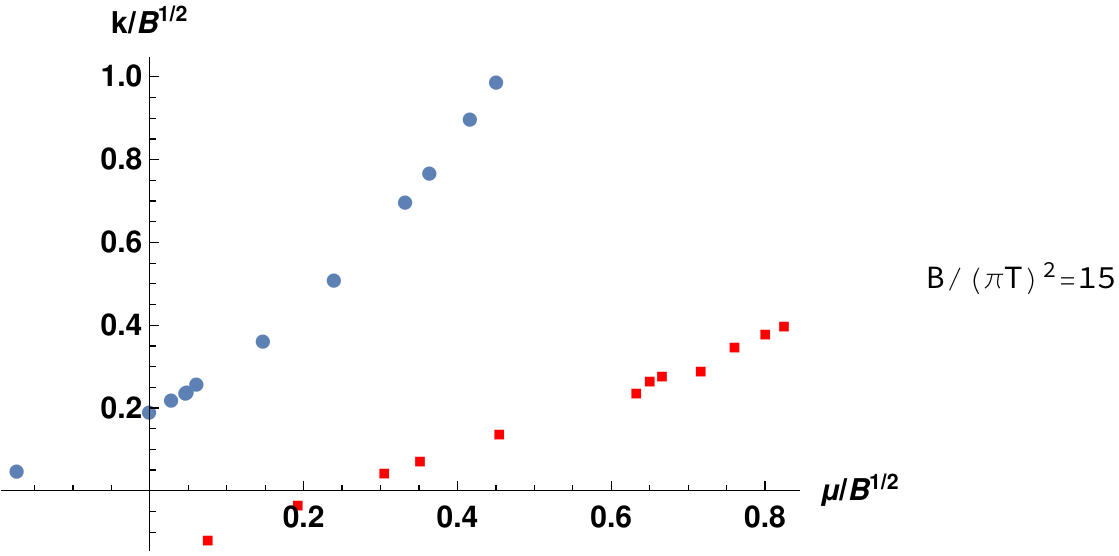}
\caption{\label{fig_b15}$n/B^{3/2}$ versus $\m/B^{1/2}$ (left) and $k/B^{1/2}$ versus $\m/B^{1/2}$ (right) at $B/(\pi T)^2=15$. The MCDW phase splits into two branches, marked by blue disks and red squares. Notably the large $k$ branch of MCDW phase extends all the way beyond $\m=0$, indicating that axial anomaly is not necessarily required for its existence. Also, the small $k$ branch extends all the way beyond $n=0(k=0)$. It is interesting to note that the behavior of $n$ and $k$ follow similar patterns.}
\end{figure}

\begin{figure}[t]
  \centering
\includegraphics[width=9cm]{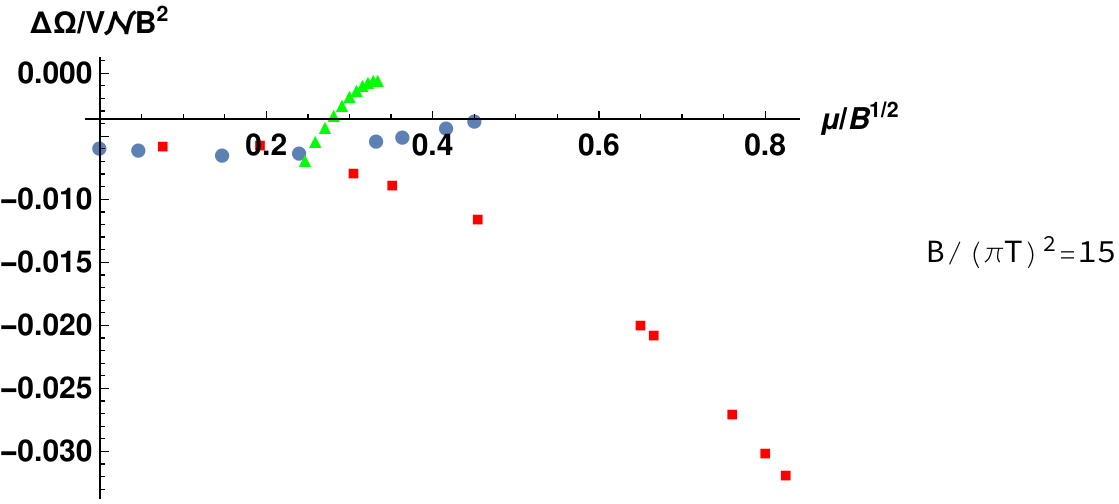}
\caption{\label{fig_b15O}$\D\O/\(V\cN B^{2}\)$ versus $\m/B^{1/2}$ at $B/(\pi T)^2=15$ for two branches of MCDW phases, marked by blue disk and red square and $\c$SB phase marked by green triangle. The small $k$ MCDW phase always has lower free energy than $\c$S phase. The large $k$ MCDW phase might be thermodynamically more favorable in region of small $\m$. The $\c$S phase exists in a narrow window of $\m$. It might be the state with the lowest free energy in an even narrower window. Current precision of numerical data does not allow for a decisive conclusion on the preferred state out of MCDW and $\c$SB phase.}
\end{figure}

Finally, we present the case of $B/(\pi T)^2=15$. In Figure \ref{fig_b15}, we show the density and spiral momentum of two branches of MCDW solutions. While the large/small density and large/small momentum correspondence still holds in general, there are also exotic cases: For large $k$ branch, the MCDW phase extends below $\m=0$, i.e. states with negative $\m$ but positive $n$ and $k$ exist. For small $k$ branch, the MCDW phase extends below $n=0(k=0)$, i.e. states with positive $\m$ but negative $n$ and $k$ exist. By continuity, we can infer that MCDW states with either $\m=0$ or $k=0$ exist.
%Note that the anomaly related WZW term is relevant for nonvanishing $\m$ and $k$.
%The presence of states with either $\m=0$ or $k=0$ suggests axial anomaly is not necessarily required for the formation MCDW phase.
We also show in Figure \ref{fig_b15O} for a comparison of free energy of different phases. The case of $B/(\pi T)^2=15$ is distinct from the cases of $B/(\pi T)^2=6.5$ and $B/(\pi T)^2=9$: the $\c$S phase is never thermodynamically preferred. In region of large $\m$, the small $k$ branch of MCDW phase is preferred. In region of small $\m$, the large $k$ branch is preferred. The $\c$SB phase exists in a narrow window in $\m$. It could be the preferred state in an even narrower window, although current precision of numerical data does not allow for a decisive answer.

\section{Anomalous Charge and MCDW Phase}\label{sec_anom}

It is interesting to discuss several aspects of the MCDW phase within the holographic model. We first discuss the role of anomalous charge. In effective models \cite{Son:2007ny}, the anomalous charge is generated from spatially inhomogeneous phase. In the presence of chemical potential, the anomalous charge can lower the free energy of the system: $\O\to\O-\m N_{\text{anom}}$. Within our holographic model, we can derive the charge density from thermodynamics
\begin{align}
  n=-\frac{\d \O}{V\d\m}=\frac{\int d\r\d{\cal L}}{\d\m}=\frac{\int d\r \d A_t'\frac{\d{\cal L}}{\d A_t'}}{\d\m}=\frac{(\d A_t(\infty)-\d A_t(1))}{\d\m}\frac{\d{\cal L}}{\d A_t'}.
\end{align}
In the last equality, we use the fact that $\frac{\d{\cal L}}{\d A_t'}$ is $\r$ independent to perform integration over $\rho$.
Note that $A_t(\infty)-A_t(1)=\m$. We thus obtain
\begin{align}
  n=\frac{\d{\cal L}}{\d A_t'}=\frac{\d{\cal L_\text{DBI}}}{\d A_t'}+\frac{\d{\cal L_\text{WZ}}}{\d A_t'}.
\end{align}
This is the conserved charge density already used in the previous section. The Lagrangian contains contribution from both DBI and WZ terms. We identify the DBI and WZ contributions as normal and anomalous charge, explicitly:
\begin{align}
  &n_{\text{norm}}=\(\cdots\)A_t',\no
  &n_{\text{anom}}=Bk(-2\c^2+\c^4).
\end{align}
Here $\(\cdots\)$ is a complicated but positive function of $A_t'$ and $\c$. In the absence of anomalous charge in homogeneous phase, it guarantees the charge density have the same sign as chemical potential. The sign of anomalous charge is instructive: note that $0<\c<1$, which gives $n_\anom>0(n_\anom<0)$ for $k>0(k<0)$. Indeed linear stability analysis \cite{Kharzeev:2011rw,Guo:2016dnm} as well as full nonlinear solution presented in this work supports positive $k$ (momentum parallel to magnetic field). This is consistent with effective model picture that formation of spiral generates anomalous charge lowering free energy of system. Had we proceeded with another gauge choice
\begin{align}\label{2c4}
C_4=\(\frac{r_0^2}{2}\r^2H\)^2dt\wg dx_1\wg dx_2\wg dx_3-\cos^4\th d\ph\wg d\O_3,
\end{align}
we would have obtained
\begin{align}
  n_{\anom}=Bk\(1-\c^2\)^2,
\end{align}
therefore $n_\anom<0(n_\anom>0)$ for $k>0(k<0)$. It implies the favorable MCDW phase should be found for $k<0$. This is not consistent with linear stability analysis and nonlinear solutions. It also serves as a confirmation of the gauge choice made in \cite{Kharzeev:2011rw} and used in this work.

Secondly, the anomalous charge defined above inherits a feature from holographic model. In effective models, normal and anomalous charge are both constant and separable, see e.g. \cite{Tatsumi:2014wka}. In holographic model, the anomalous charge, as well as the normal charge depends on holographic coordinate $\r$. Only the sum of the two is a constant. It is known that the holographic coordinate plays the role of renormalization group (RG) scale. It is interesting to analyze the variation of $n_\anom$ along RG scale: since $\c=0$ at both horizon and boundary, we conclude that $n_\anom$ vanishes in the IR and UV limits. In intermediate scale, $n_\anom>0$. To construct an effective model based on holographic theory, we would need to integrate out the holographic coordinate from UV to certain cutoff scale in the middle. The resultant effective anomalous charge is not expected to be a simple product $Bk$, in contrast to effective models.

\begin{figure}[t]
  \centering
\includegraphics[width=9cm]{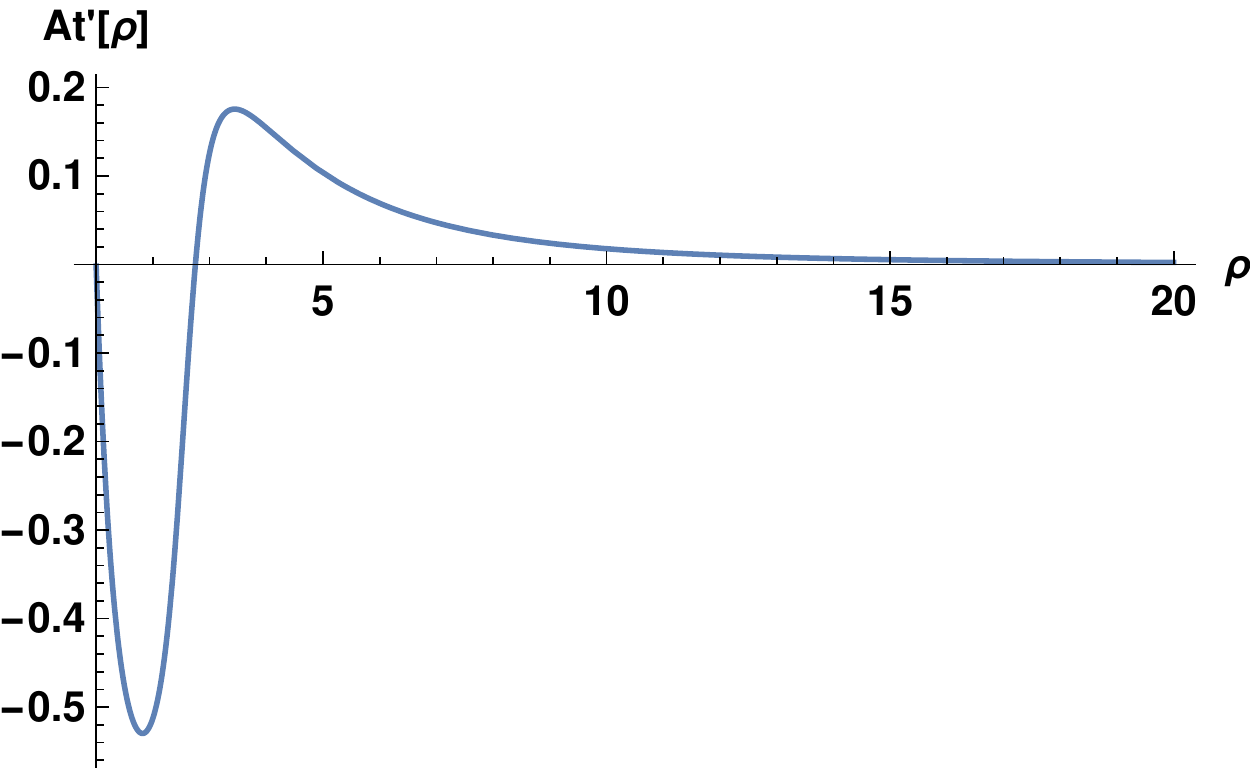}
\caption{\label{fig_atp}$A_t'(\r)$ at $B/(\pi T)^2=15$. The positive and negative contributions in $\int d\r A_t'(\r)$ cancel out giving a vanishing $\m$. There is one zero of $A_t'(\r)$, at which $n_{\text{norm}}=0$ and $n_{\anom}=Bk\(2\c^2-\c^4\)$. This explains why $n$ and $k$ have the same sign.}
\end{figure}

Finally, we discuss the two exotic MCDW states at $B/(\pi T)^2=15$ and their relation with axial anomaly. One state has $\m=0$, but $n\ne0(k\ne0)$. According to the definition \eqref{mu_def}, $A_t'$ has at least one zero. We confirm this by plotting $A_t'(\r)$ in Figure \ref{fig_atp}.
Naively axial anomaly is not relevant for $\m=0$. This is not true: although the integration of $A_t'(\r)$ vanishes, the integration of WZ term is non-vanishing, which contributes to the thermodynamics. Mathematically, the contributions from DBI and WZ terms take the following form
\begin{align}
  &\O_{\text{DBI}}^n/V\ne-\int d\r A_t'n_{\text{norm}},
  &\O_{\text{WZ}}^n/V=-\int d\r A_t'n_{\anom}.
\end{align}
We use the superscript $n$ to indicate that they are contribution from density.
The WZ term is a simple coupling between chemical potential and $n_\anom$, while the DBI term cannot be written as a simple coupling between chemical potential and $n_{\text{norm}}$ due to the nonlinear dependence of DBI action on $A_t'$. If this were true, we could combine the two terms by using $n_{\text{norm}}+n_\anom=\text{constant}$, giving a vanishing contribution because $\m=\int d\r A_t'=0$. However due to different nature of anomalous charge and normal charge, anomaly can still play a role even at $\m=0$.

\begin{figure}[t]
  \centering
\includegraphics[width=9cm]{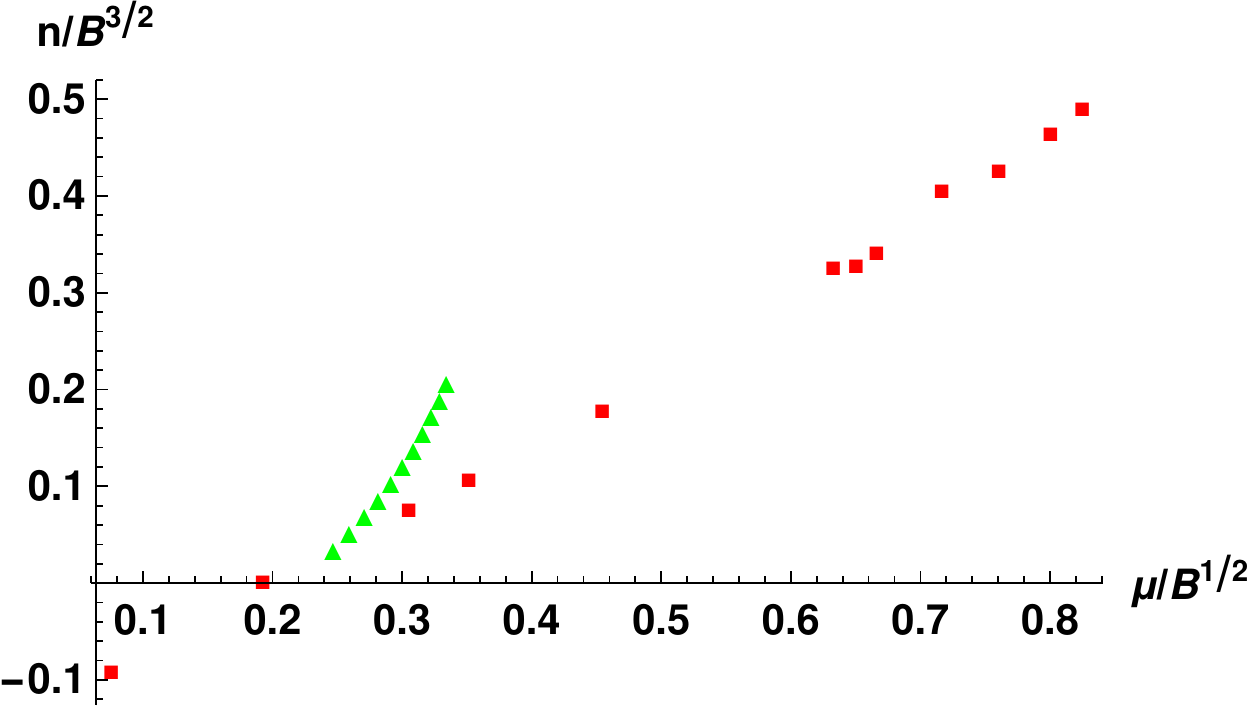}
\caption{\label{fig_pt}$n/B^{3/2}$ versus $\m/B^{1/2}$ at $B/(\pi T)^2=15$ for small $k$ branch of MCDW phase (red squares) and $\c$SB phase (green triangles). At $\m/B^{1/2}\simeq 0.25$, the density corresponding to two phases merge, suggesting a second order phase transition. The critical value of $\m$ agrees with the $k=0$ state of MCDW phase in Fig~\ref{fig_b15} and also the free energy comparison in Fig~\ref{fig_b15O}.}
\end{figure}

The other two states have $n=0$ and $k=0$ respectively. Although they lie close in $\m$ numerically, we can argue they are different states. For state with $n=0$, we need $n_{\text{norm}}$ and $n_\anom$ to cancel each other. Since $n_{\text{norm}}$ is in general nonvanishing for arbitrary $\r$, $n_\anom$ must also be nonvanishing. Thus we cannot have a state with $n=0$ and $k=0$ simultaneously. The state with $n=0$ and $k\ne0$ is still related to axial anomaly as we need anomalous charge to cancel normal charge. The state with $k=0$ and $n\ne0$ is homogeneous, thus it should reduce to the $\c$SB case. In Figure \ref{fig_pt} we show a comparison of density and chiral condensate between MCDW phase and $\c$SB phase. It confirms a continuous merging of the two phases. Combining with Fig.~\ref{fig_b15O}, we suggest that the $\c$SB phase may be replaced by MCDW phase.

\section{Summary and Outlook}\label{sec_sum}

We explore the end point of the spiral instability studied in \cite{Kharzeev:2011rw}. We find the end point solution contains both chiral condensate and pseudoscalar condensate, analogous to magnetized chiral density wave phase in literature \cite{Tatsumi:2014wka}. The MCDW phase contains two branches of solutions, in accordance with the number of unstable modes found in \cite{Kharzeev:2011rw,Guo:2016dnm}. Within each branch, the momentum $k$ can take continuous values. Minimizing the free energy with respect to $k$ gives the thermodynamically preferred state. We find for not large $B$, the large $k$ branch of the MCDW phase is the preferred state out of the two branches. In this case, there is a critical $\m$, beyond which the MCDW phase dominates over $\c$S and $\c$SB phases.
For large $B$, the small $k$ branch becomes preferred out of the two branches for wide range of $\m$. At sufficient large $\m$, the MCDW phase becomes dominant over $\c$S and $\c$SB phases.

We also give a holographic definition of anomalous charge. The anomalous charge in holographic model varies along RG flow. In particular, it vanishes in the IR and UV limits in our model, but is finite in the intermediate scale. The sum of anomalous and normal charge is constant along the RG flow.

We also find an exotic state of MCDW phase at large $B$ and vanishing $\m$. Surprisingly axial anomaly still plays a role at vanishing $\m$, leading to formation of spiral phase. The reason is normal charge and anomalous charge respond to $\m$ differently. The free energy can be lowered by forming nonvanishing sum of the two.

This work can be extended in a few directions. First of all, we focus on finite density states in this work. To have a complete study of phase diagram, we still need zero density states. The homogeneous zero density states have been studied in \cite{Evans:2010iy}. It would be interesting to see whether MCDW phase exists at zero density. A closely related question is to find out whether magnetized kink solution can be realized in holographic models and how it may change the phase diagram.

Secondly, at strong magnetic field and finite $\m$ or finite axial chemical potential $\m_5$, the ground state is conjectured to be chiral magnetic spiral phase. Unlike longitudinal spiral (along magnetic field), it is featured by transverse spiral. While the case with $\m_5\ne0$ is confirmed in holographic model study \cite{Kim:2010pu}, the case with $\m\ne0$ is not found in the same study. It is desirable to have an independent check within our model.

Last but not least, it would also be interesting to explore the transports of MCDW phase. Since MCDW phase breaks both chiral symmetry and translational symmetry, it would be interesting to study the corresponding Nambu-Goldstone modes, and moreover the hydrodynamics in MCDW phase background. We leave these for future studies.

\section*{Acknowledgments}

S.L. is grateful to Gaoqing Cao, Yoshimasa Hidaka and Keun-Young Kim for useful discussions. S.L. is supported by One Thousand Talent Program for Young Scholars and NSFC under Grant Nos 11675274 and 11735007. Y.B. is supported by the Fundamental Research Funds for the Central Universities under grant No.122050205032 and the NSFC under the grant No.11705037.

\appendix

\bibliographystyle{unsrt}
\bibliography{Q5ref}

\end{document}